\documentclass[reprint,superscriptaddress,showpacs,amsmath,amssymb,aps,prl]{revtex4-1}

\usepackage[usenames]{color}
\usepackage{graphicx}
\usepackage{dcolumn}
\usepackage{bm}

\begin{document}
\title{Measurement of the Target-Normal Single-Spin Asymmetry in Deep-Inelastic Scattering 
from the Reaction $^{3}\mathrm{He}^{\uparrow}(e,e^\prime)X$}   


\author{J. Katich} 
\affiliation{College of William and Mary, Williamsburg, VA 23187}
\affiliation{University of Colorado, Boulder, CO 80309}
\author{X.~Qian} 
\affiliation{Duke University, Durham, NC 27708}
\affiliation{Kellogg Radiation Laboratory, California Institute of Technology, Pasadena, CA, 91125}
\affiliation{Brookhaven National Laboratory, Upton, NY 11973}
\author{Y. X. Zhao}
\affiliation{University of Science and Technology of China, Hefei 230026, People's Republic of China}
\author{K.~Allada}
\affiliation{University of Kentucky, Lexington, KY 40506}
\author{K.~Aniol}
\affiliation{California State University, Los Angeles, Los Angeles, CA 90032}
\author{J. R. M.~Annand}
\affiliation{University of Glasgow, Glasgow G12 8QQ, Scotland, United Kingdom}
\author{T.~Averett}\email[Corresponding author: ]{tdaver@wm.edu}
\affiliation{College of William and Mary, Williamsburg, VA 23187}
\author{F.~Benmokhtar}
\affiliation{Carnegie Mellon University, Pittsburgh, PA 15213}
\author{W.~Bertozzi}
\affiliation{Massachusetts Institute of Technology, Cambridge, MA 02139}
\author{P.C.~Bradshaw}
\affiliation{College of William and Mary, Williamsburg, VA 23187}
% ...

\date{\today}

\begin{abstract}
We report the first measurement of the target-normal single-spin asymmetry in deep-inelastic scattering from the inclusive reaction $^3$He$^{\uparrow}\left(e,e' \right)X$  on a polarized $^3$He gas target. 
Assuming time-reversal invariance, this asymmetry is strictly zero in the Born approximation but can be non-zero if two-photon-exchange contributions are included.  The experiment, conducted at Jefferson Lab 
using a 5.89 GeV electron beam, covers a range of $1.7 < W < 2.9$ GeV, $1.0<Q^2<4.0$ GeV$^2$ 
and $0.16<x<0.65$.  Neutron asymmetries 
were extracted using the effective nucleon polarization and measured proton-to-$^3$He cross 
section ratios. The measured neutron asymmetries are negative with an average value of $(-1.09 \pm 0.38) \times10^{-2}$ for invariant mass $W>2$ GeV, which is non-zero 
at the $2.89\sigma$ level.   Our  measured asymmetry  agrees both in sign and magnitude with a two-photon-exchange model prediction that uses input  from the Sivers transverse momentum distribution obtained from semi-inclusive deep-inelastic scattering.
\end{abstract}


\maketitle

\begin{figure}[h]
\includegraphics[width=3in]{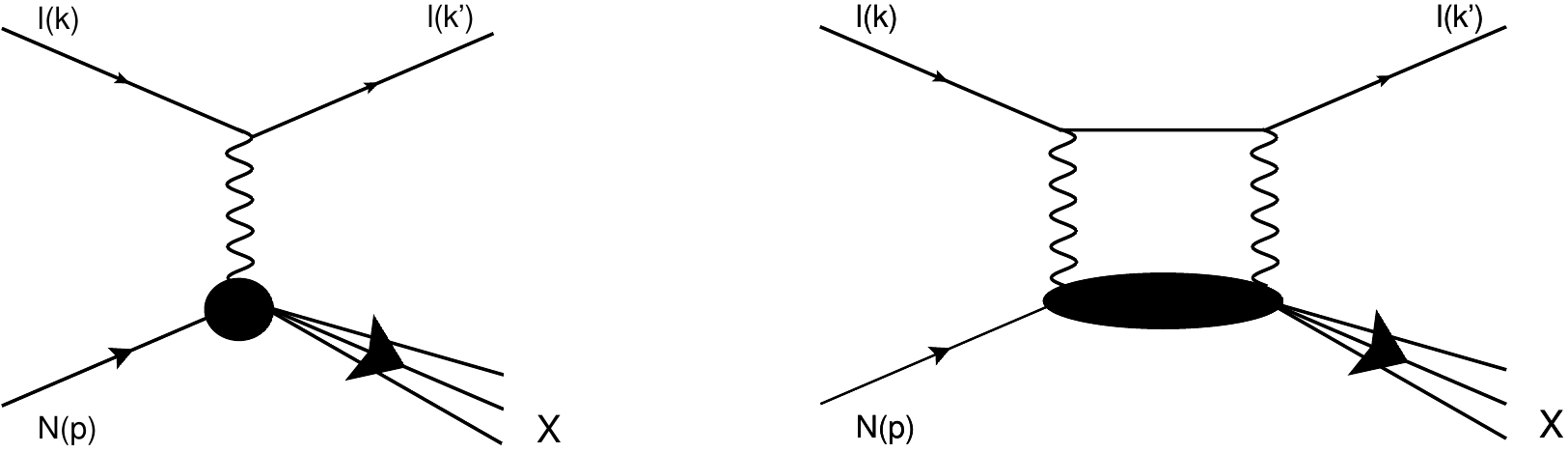}
\caption{Interference between one- and two-photon exchange in N$(e,e')$ allows the possibility of a non-zero target SSA. Here, $l$ is 
the lepton  with incident and outgoing 4-momenta $k$ and $k'$, respectively. N is 
the nucleon with initial 4-momentum  $p$.} 
\label{fig:TPEX_feynman} 
\end{figure}

The past decade has seen a resurrection of interest in two-photon exchange in 
electron-nucleon scattering.  This is primarily due to the realization that inclusion 
of the two-photon-exchange amplitude can partially reconcile the discrepancy between 
the Rosenbluth separation and the polarization-transfer methods for extracting the $Q^2$-dependence of the proton 
elastic form factor ratio, $G^p_E/G^p_M$~\cite{g63, Arrington:2011dn, Christy:2004rc, Qattan:2004ht, Puckett:2011xg, Blunden_GeGm,  Chen:2004tw, carl_2gamma}.  As the precision of nucleon structure measurements improves, it is important to 
understand the dynamics of the two-photon-exchange processes.  Assuming conservation of parity and time-reversal invariance, the target single-spin asymmetry (SSA) in $(e,e')$  from a target polarized normal  to the electron scattering plane is strictly zero at Born level~\cite{Christ_Lee}, but can be non-zero when  interference between one- and two-photon exchange processes is included (Fig.~\ref{fig:TPEX_feynman}).  

\begin{figure}[h]
\includegraphics[width=3in]{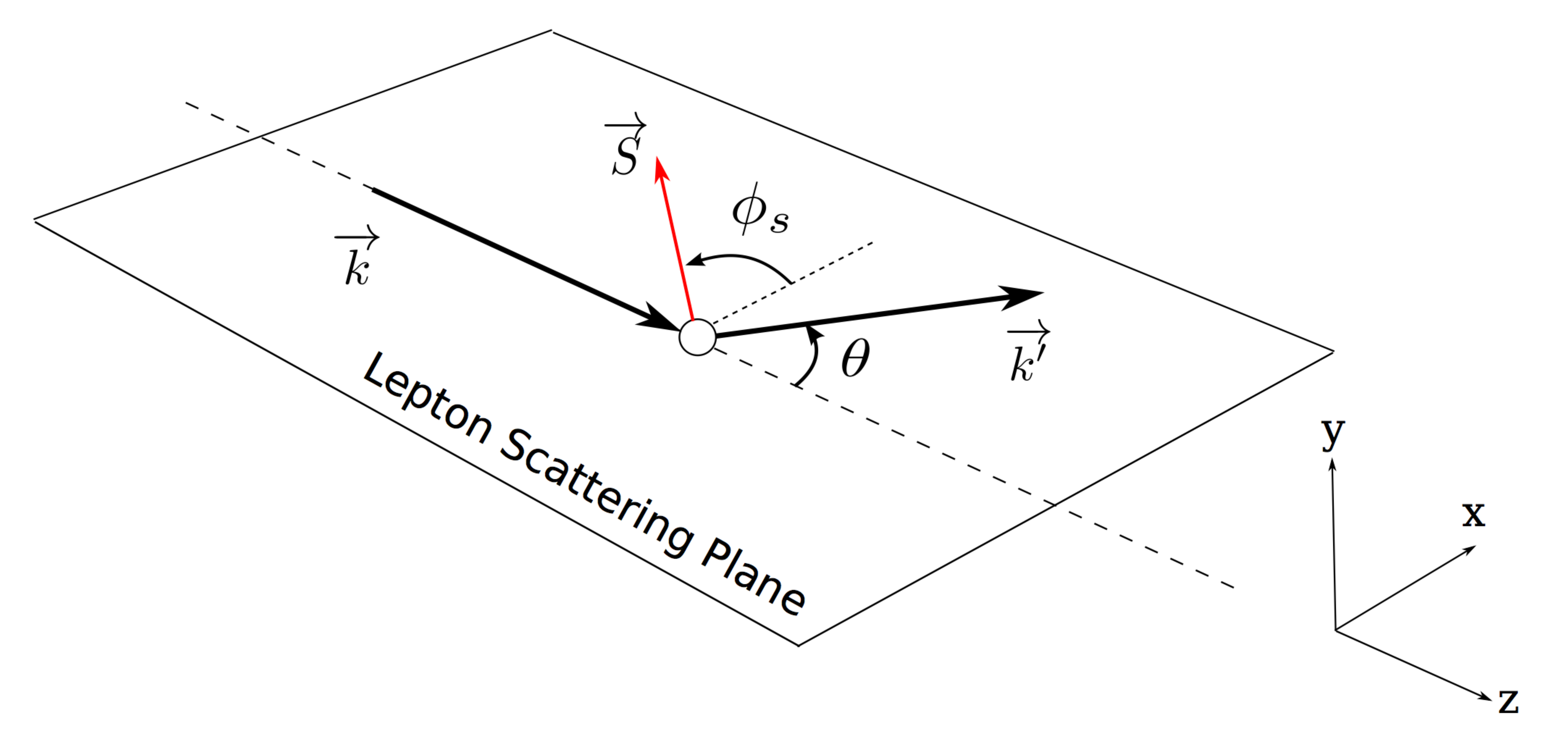}
\caption{Coordinate system used to define $A_{UT}(\phi_S)$.} 
\label{fig:coords} 
\end{figure}

Consider the inelastic scattering of an unpolarized electron from a target nucleon with vector 
spin $\vec{S}$, oriented perpendicular (transversely polarized) to the incident electron 3-momentum $\vec{k}$, with normalization $|\vec{S}|=1$.   Requiring conservation of the electromagnetic current 
and parity, the differential cross section, $d\sigma$, for inclusive scattering is 
written as~\cite{Christ_Lee, Cahn_2g, Afanasev_DIS} 
\begin{eqnarray}
d\sigma(\phi_S)  &= d\sigma_{UU} + \frac{{\vec{S}}\cdot({\vec{k}} \times {\vec{k'}})}{|{\vec{k}} \times {\vec{k'}}|}   d\sigma_{UT}\\
&= d\sigma_{UU} + d\sigma_{UT} \sin \phi_S,  
\end{eqnarray}
where $\vec{k'}$ is the  3-momentum of the scattered electron, and  $d\sigma_{UU}$ and $d\sigma_{UT}$ are the cross sections for an unpolarized electron scattered from an unpolarized and transversely polarized target, respectively.   
Our choice of coordinates is shown in Fig.~\ref{fig:coords} with the angle $\phi_S$ between the lepton plane and $\vec{S}$.  The $+\hat{y}$ direction is parallel to the vector ${\vec{k}} \times {\vec{k'}}$ and corresponds to $\phi_s=90^{\circ}$. 
We define the SSA as
\begin{equation}\label{eq:ay}
A_{UT} (\phi_S) = \frac{d\sigma(\phi_S)-d\sigma(\phi_S + \pi)}{d\sigma(\phi_S)+d\sigma(\phi_S + \pi)}
 =  A_y \sin \phi_S.
\end{equation}
The quantity $A_y \equiv \frac{d\sigma_{UT}}{d\sigma_{UU}}$ can be extracted by measuring the $\phi_S$-dependence of  $A_{UT}(\phi_S)$, or by measuring the SSA for a target 
polarized normal to the lepton plane.

Considering only the one-photon-exchange amplitude, $\mathcal{M}_{1\gamma}$, we can write 
$d\sigma_{UU} \propto \mathcal{R}e(\mathcal{M}_{1\gamma} \mathcal{M}_{1\gamma}^*)$ and 
$d\sigma_{UT}\propto \mathcal{I}m(\mathcal{M}_{1\gamma} \mathcal{M}_{1\gamma}^*)$, where 
$\mathcal{R}e$ ($\mathcal{I}m$) stands for the real (imaginary) part.  However time-reversal 
invariance  requires that $\mathcal{M}_{1\gamma}$ be real and so at order $\alpha_{em}^2$, $d\sigma_{UU}$ 
can be non-zero but $d\sigma_{UT}$ must be zero.  When one includes the (complex) two-photon-exchange 
amplitude, $\mathcal{M}_{2\gamma}$, the contribution to  the asymmetry from  one- and two-photon 
interference is $d\sigma_{UT}\propto \mathcal{I}m(\mathcal{M}_{1\gamma} \mathcal{M}_{2\gamma}^*)$ 
which can be non-zero at order $\alpha_{em}^3$. The two-photon exchange process forms a loop with the nucleon intermediate state and contains the full response of the nucleon (see Fig.~\ref{fig:TPEX_feynman}).  

An additional contribution to $d\sigma_{UT}$ at order $\alpha_{em}^3$ may  arise from  interference 
between real photon emission (bremsstrahlung) by the electron and the hadronic system.  Detailed discussions of these contributions are presented in Refs.~\cite{Metz, Schlegel:2012ve, Afanasev_DIS}.

There are no published measurements of $A_y$ for the neutron.  
For protons, the first measurement of $A_y^p$ was done in 1968 at CEA~\cite{CEA}.  
Electrons were scattered from an alcohol/water target containing  
protons with an average polarization $\sim 20\%$.  Three invariant photon-hadron masses 
were studied,  $W=$1236, 1512 and 1688 MeV, with $Q^2=0.2 - 0.7$  GeV$^2$.   
Results were consistent with zero at the $4 \times$ $10^{-2}$ level.
In 1969 a  measurement  at SLAC~\cite{SLAC} was made using both $e^-$ and $e^+$ scattering in the resonance region with $Q^2=0.4-1.0$ GeV$^2$.  A butanol target provided protons with a polarization of $\sim20\%$.  Results were consistent with zero at the  few $\times 10^{-2}$ level.

A theoretical calculation for $A_y^p$ at $W=1232$ MeV~\cite{Cahn_2g} 
treated the intermediate state as purely elastic and predicted   
$A_y^p \sim 0.75 \times 10^{-2}$ at $Q^2=0.6$ GeV$^2$.

The only measurement of $A_y^p$ using deep-inelastic scattering (DIS)
was made at DESY by the HERMES collaboration~\cite{HERMES}.
Both $e^-$ and $e^+$ with energy 27.6 GeV  were scattered from a  polarized  hydrogen target with average polarization $\sim 75\%$.  
Particles were detected over $0.007 < x_B < 0.9$, $0.25 < Q^2 < 20$~GeV$^2$ and $\phi_S= 0 - 2\pi$.  Results for $A_y^p$  for Q$^2 > 1$~GeV$^2$ are consistent with zero  at the $\sim10^{-3}$ level.

There are two parton-model predictions for the two-photon exchange contribution to $A_y$ for protons and neutrons in DIS.  The first, by  A.\  Afanasev {\it et al.}~\cite{Afanasev_DIS} assumes 
the scattering is dominated by two-photon exchange with a single quark and predicts $A_y^n\sim10^{-4}$  
at $x\sim0.3$ and $Q^2=2.0$ GeV$^2$.  In the second prediction,  A. Metz {\it et al.}~\cite{Metz} assume the asymmetry is dominated 
by the process where one of the photons couples to an active quark and the other couples to a quark in the spectator di-quark system.    When the interaction with the di-quark system is modeled using input from the Sivers distributions from semi-inclusive DIS~\cite{Alekseev:2008aa,Airapetian:2009ae}, they predict $A_y^n \sim -10^{-2}$ at
the kinematics of our experiment.   For consistency with our sign convention, the asymmetries in Ref.~\cite{Metz} have been multiplied by $-1$.

This paper presents the results of Jefferson Lab experiment E07-013, which was a 
measurement of the neutron SSA, $A_y^n$, in DIS.  The $\phi_S$-dependent 
asymmetries were measured using inclusive scattering of unpolarized electrons from a 
$^3$He target polarized either vertically ($\phi_S \sim \pm 90^{\circ}$)
or transversely ($\phi_S \sim 0^{\circ},~180^{\circ}$) in the lab frame.
$A_y$ was obtained by fitting the $\phi_S$ dependence according to 
Eqn.~\eqref{eq:ay}.  The nuclear ground state of $^3$He 
is dominated by the configuration in which the spins of two protons are anti-aligned, which means 
that the spin is mostly carried by the neutron, effectively providing a polarized 
neutron target. 

An electron beam with  energy  $5.889$ GeV and  average current  $12$ $\mu$A was incident on polarized $^3$He gas with density  $\sim10$~amg contained in a  $40$~cm-long cylindrical aluminosilicate glass cell. 
The beam was rastered in a $3\times3$~mm$^2$ pattern to reduce the possibility 
of cell rupture and localized de-polarization.  Polarization of the $^3$He nuclei was achieved via 
Spin-Exchange Optical Pumping (SEOP) with a hybrid alkali-metal mixture of Rb and K~\cite{Babcock03}.  
The polarization direction was reversed every $20$ minutes using adiabatic fast passage nuclear 
magnetic resonance (NMR).  With each spin-flip, the NMR signals were used to measure the relative 
polarization.  Absolute calibration was done periodically throughout the run using electron 
paramagnetic resonance~\cite{Romalis_EPR}.  The average polarization was  $55\%$ with a 5\% relative
 uncertainty.  The total luminosity downstream of the target was measured during each 20-minute target 
polarization state using eight Lucite/PMT detectors placed symmetrically around the beam line.  
The average luminosity asymmetry for the experiment was $(38\pm 12 )\times 10^{-6}$  
which is negligible compared to our measured raw asymmetries of $\sim 10^{-3}$.

Scattered electrons were detected using the  Hall A BigBite detector package~\cite{Mihovilovic:2012hi} 
at $+30^\circ$ (beam-right) and the left Hall A High Resolution Spectrometer 
(LHRS) at $-16^\circ$ ~\cite{halla_nim}.  The BigBite package includes a dipole magnet for momentum separation, three sets of multi-wire drift chambers 
for track reconstruction, and a lead-glass electromagnetic calorimeter for particle identification (PID) with pre-shower and shower layers sandwiching a scintillator plane for providing timing 
information.  The useful momentum coverage of BigBite was $0.6 < p < 2.5$ GeV with an average 
solid angle acceptance of 64 msr. The corresponding $\phi_S$ coverage is $\sim 60^{\circ}$ for each target polarization configuration.
The LHRS consists of two sets of drift chambers for tracking,
 two scintillator planes for the trigger, and  gas Cherenkov  and lead-glass shower 
detectors for  PID.  The central momentum of the LHRS was $2.35$ GeV
with a momentum coverage of $\pm$ 4.5\%.  The solid angle acceptance was  $\sim 6$~msr 
with  $\sim 7^{\circ}$ $\phi_S$ coverage. 
Optics for both detectors were calibrated using elastic $e^-$ scattering from 
hydrogen and multi-foil carbon targets. Angular reconstruction in both
detectors was calibrated using a sieve slit placed in front of each spectrometer. 
The angular resolution in BigBite was  $<10$ mrad and the the resolution of the reconstructed 
momentum was  $<1\%$.  

\begin{table*}[ht]
\centering
\begin{tabular}{|c|c|c|c|c|c|c|}
\hline
Detector & $W$     & $x$    & $Q^2$     & $A_y^{^3\mathrm{He}} \pm$ (stat)  $\pm$ (sys)      & $A_y^n\pm $ (stat)  $\pm$ (sys)  & Pair-produced background     \\
         & GeV   &        & GeV$^2$   & $(\times 10^{-3})$   &  $(\times 10^{-2})$   & contamination (\%)\\
\hline
BigBite &  1.72   & 0.65   & 3.98   & $-0.85  \pm 2.79  \pm 0.53$  & $-0.55	  \pm  1.81	      \pm 	  0.36$   & $1.0 \pm 0.8$ \\               
BigBite &  2.17   & 0.46   & 3.24   & $-6.28  \pm 2.51  \pm 0.88$  & $-3.87       \pm  1.55           \pm         0.58$   & $3.1 \pm 1.1$\\               
BigBite &  2.46   & 0.34   & 2.65   & $-8.14  \pm 1.99  \pm 1.05$  & $-3.89       \pm  0.96           \pm         0.53$   & $9.5 \pm 2.0$\\               
BigBite &  2.70   & 0.24   & 2.08   & $-2.25  \pm 2.45  \pm 1.46$  & $-1.08       \pm  1.18           \pm         0.69$   & $22.0 \pm 4.5$\\               
BigBite &  2.89   & 0.17   & 1.58   & $-8.34  \pm 4.35  \pm 5.33$  & $-3.84       \pm  2.00           \pm         2.42$   & $48 \pm 10$\\               
\hline
LHRS    &  2.54  &  0.16  &  1.05  & $-1.57  \pm 0.99 \pm 0.2$ & $-0.64      \pm  0.41           \pm         0.09$   & $1.3 \pm 0.05$\\               
\hline
\end{tabular}
\centering
  \caption{Kinematics and results for neutron asymmetries with 
statistical and systematic uncertainties.  The BigBite spectrometer was set at a 
fixed angle and central momentum and data were divided into the five kinematic bins.  The final column shows measured contaminations from pair-produced electrons.}
\label{table:kine}
\end{table*}

Electron PID in BigBite began at the trigger level, which 
required the sum of the pre-shower and shower signals to be above a chosen threshold.   
Events with poor track reconstruction, tracks near the edges of the acceptance, and data that could be affected by beam trips were removed.  Additional cuts included particle charge, 
reconstructed particle momentum, reconstructed vertex, energy deposited in the pre-shower 
detector ($E_{ps}>$ 200 MeV), and a cut on the ratio of reconstructed energy to reconstructed 
momentum ($E/p$).  The LHRS cuts were similar and included cuts on the reconstructed vertex, 
Cherenkov amplitude, and an $E/p$ cut. The data from BigBite covered $0.17 < x < 0.65$ and were divided 
into five bins in $W$.  The LHRS data was analyzed as a single kinematic point ($x = 0.16$, $W=2.54$ GeV).

Events from three triggers taken simultaneously were used in the BigBite analysis. 
They are T1, proportional to the total energy deposited in the
calorimeter, T6, which is the same as T1 but with higher discriminator threshold,
and T2, coincidence between a gas Cherenkov detector
and T6. Prescale factors ranging from 2100 to 3100, 61 to 410,
 and 350 to 780 were used for T1, T2 and T6, respectively.
Because the background rate from the Cherenkov 
detector was extremely high, the T2 trigger is functionally the same as the T6 trigger. 
Information from the Cherenkov detector was not used in this analysis. 
In the final dataset, T6 contributes to more than 80\% of the 
data while T2 is about 12\% and T1 is less than 8\%. 

Raw asymmetries for each data bin were formed as
\begin{equation}
 A^{e^-}_{raw} (\phi_S) = \frac{1}{P_{target}} \frac {Y^{\uparrow}_{raw} (\phi_S) - Y^{\downarrow}_{raw} (\phi_S + \pi)}  {Y^{\uparrow}_{raw}(\phi_S) + Y^{\downarrow}_{raw}(\phi_S + \pi)}
\end{equation}
\noindent where the raw  yields, $Y^{\uparrow(\downarrow)}_{raw}$, are the number of particles, 
$N$, observed in the target spin ``up" (``down") state that pass all data cuts for electrons, 
normalized by accumulated charge, $Q$, and DAQ livetime, $LT$:
\begin{equation}\label{eq:yield}
 Y^{\uparrow (\downarrow)}_{raw} = \frac{N^{\uparrow (\downarrow)}_{raw}} {Q^{\uparrow (\downarrow)} \cdot LT^{\uparrow (\downarrow)}}  = \frac{     N^{\uparrow (\downarrow)}_{e^-} +  N^{\uparrow (\downarrow)}_{\pi^-}   +  N^{\uparrow (\downarrow)}_{e^+}     } {Q^{\uparrow (\downarrow)} \cdot LT^{\uparrow (\downarrow)}}.
 \end{equation}
 The terms $N_{\pi-}$ and $N_{e+}$ represent  pion and pair-produced electron backgrounds that pass the good-electron cuts and $P_{target}$ 
is the target polarization.  The $\phi_S$ angle is defined for the spin up state, and 
changed by $180^{\circ}$ ($\phi_S + \pi$) when the target spin was flipped.

The dominant background passing the data cuts in BigBite were photo-induced electron-positron 
pairs.  The positrons were cut from the data by requiring  particles with negative charge.  
However, the pair-produced electrons are indistinguishable from the desired DIS electrons.  
A direct measurement of the pair-produced electron contamination was made by reversing the 
polarity of the BigBite magnet and calculating the positron yield under conditions identical 
to the normal data collection.  Since photons are mostly produced from 
neutral pion decay, the contamination decreased with increasing momentum, see Table~\ref{table:kine}.
This also explains why this type of background in the LHRS (central momentum of 2.35 GeV) is negligible.
Negative pions were also a source of contamination.  Their 
contributions to the BigBite data   were  accounted for by fitting the pre-shower energy  spectrum.  
Likewise, the positron data sample was contaminated by positive pions. The positive pion contamination
was estimated based on the negative pion contamination. A GEANT-based Monte Carlo simulation of the BigBite
spectrometer was used to study the differences between the $\pi^+$ and $\pi^-$ contaminations. 
Data from the LHRS were relatively free of background contamination due to the choice of kinematics and 
exceptional PID.

Due to the large acceptance of the BigBite spectrometer, asymmetries for each type of background particle 
($A^{\pi^-}$, $A^{e^+}_{raw}$, and $A^{\pi^+}$) were obtained from the data in the same way as $A^{e^-}_{raw}$
but with different selection cuts: i)  positrons were selected using the same cuts as electrons except
for the particle charge; ii)  pions were selected using the same cuts as electrons/positrons 
except for requiring  a pre-shower energy deposition  under $150$ MeV.
Corrections were made to the asymmetry via:
\begin{equation}
A^{e^-} = \frac{A^{e^-}_{raw} - f_{1} A^{\pi^-} - f_4\left(1-f_3\right)\frac{A^{e^+}_{raw}-f_5A^{\pi^+}}{1-f_5}}{1 - f_1 - f_4\left(1-f_3\right)},
\label{eqn:corrections}
\end{equation}
where the coefficients, $f_i$, give the fractions of mis-identified particles and are defined as:
\begin{eqnarray}
\nonumber f_1=Y^{\pi^-}_{neg} / (Y^{e^-}_{neg}+Y^{\pi^-}_{neg})~~~~~~~\\ 
\nonumber f_3=Y^{\pi^+}_{pos} / (Y^{e^+}_{pos}+Y^{\pi^+}_{pos})~~~~~~~~\\
\nonumber f_4=(Y^{e^+}_{pos}+Y^{\pi^+}_{pos} ) / (Y^{e^-}_{neg}+Y^{\pi^-}_{neg})\\
f_5=Y^{\pi^+}_{neg} / (Y^{e^+}_{neg}+Y^{\pi^+}_{neg}).~~~~~~
\label{eqn:correction_input}
\end{eqnarray}
The $pos$ and $neg$ subscripts indicate the polarity of the BigBite magnet 
(standard running conditions are $neg$). The $f_5$ were estimated based on $f_3$.  Further information on these background 
corrections is provided in the appendix.

A small quantity of unpolarized N$_2$ was used in the $^3$He target-cell to improve the 
efficiency of the optical pumping.  The asymmetry was corrected by a dilution factor defined as:
\begin{equation}
\eta_{\mathrm{N}_2} \equiv \frac{1}{1+\left(\frac{\rho_{\mathrm{N}_2}}{\rho_{^3\mathrm{He}}}\right)\left(\frac{\sigma_{\mathrm{N}_2}}{\sigma_{^3\mathrm{He}}}\right)}
\end{equation}
 where $\rho$ are the densities and $\sigma$ are the unpolarized cross-sections for each gas.  The ratio 
of densities is taken from the target cell filling data.  The cross-section ratio is determined 
experimentally by inelastic scattering from a reference cell filled with known densities 
of either N$_2$ or $^3$He. 
The dilution factors for BigBite measured for T1 and T6 triggers
agree with each other. The final dilution was determined by 
combining results from T1 and T6 according to their statistical uncertainties, 
giving $\eta \sim 0.9$ for all kinematics
with an uncertainty of $\sim 2\%$. 
The dilution factor for the LHRS was determined to be 0.851 $\pm$ 0.018.
The $^3$He asymmetries from BigBite T1, T2 and T6 triggers 
were extracted independently and were consistent with each other within the statistical uncertainties for each bin.  The final $^3$He asymmetries were obtained by combining the results 
from the T1, T2 and T6 asymmetries according to their statistical 
uncertainties.

Neutron asymmetries were obtained from the $^3$He asymmetries using the effective 
polarizations of the proton and neutron in polarized $^3$He using~\cite{Scopetta},
\begin{equation}
A_y^{^3\mathrm{He}}=(1-f_p) P_n A_y^n + f_p P_p A_y^p
\label{eq:3he_n}
\end{equation}
Here,  $P_n=0.86^{+0.036}_{-0.02}$ ($P_p=-0.028^{+0.009}_{-0.004}$) is the effective neutron
(proton)  polarization~\cite{Zheng_PRC}.

The proton dilutions of $^3$He for BigBite, $f_{p} = \frac{2\sigma_p}{\sigma_{\rm ^3He}}$,  
were measured for the T1 and T6 triggers using the yields from unpolarized 
hydrogen and $^3$He targets and are consistent with each other. 
The final dilutions, which varied between $0.75-0.82$, with uncertainties of $0.02-0.08$, were determined by combining the T1 and T6 results according to their statistical uncertainties.  Neutron asymmetries were calculated separately for each trigger type and combined according to their statistical uncertainties.
The proton dilution for the LHRS was 
0.715 $\pm$ 0.007.  A value of $A_y^p=(0 \pm 3)\times 10^{-3}$ was used in Eqn.~\eqref{eq:3he_n} based on the HERMES measurements~\cite{HERMES}.  
External radiative corrections were applied to both the BigBite and LHRS data using a Monte Carlo simulation that 
included detailed modeling of geometry and material in the target and spectrometers. No correction was made on the asymmetries since the radiative
corrections to the two-photon exchange process are not yet 
available and the phase space of this measurement is limited.
\begin{figure}[h]
\includegraphics[width=3.5in, height=3.0in]{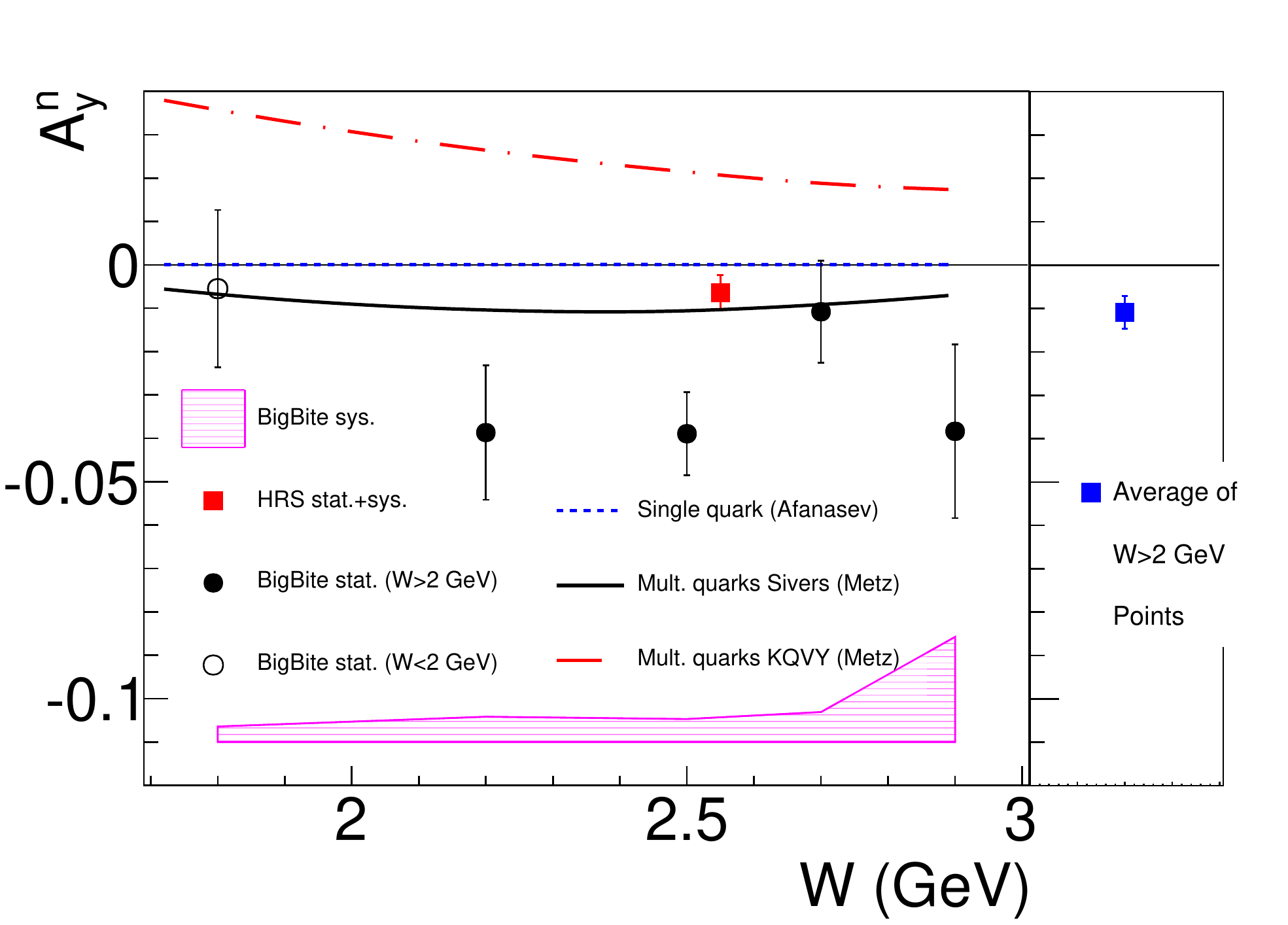}
\caption{\label{fig:DIS_Ay_n} Neutron asymmetry results (color online). {\bf Left panel:}  Solid black data points are DIS data ($W>2$ GeV) from the BigBite spectrometer; open circle has $W=1.72$ GeV.  BigBite data points show statistical uncertainties with systematic uncertainties indicated by the lower solid band.
 The square point is the LHRS data with combined statistical and systematic uncertainties.  The dotted curve near zero (positive)
is the calculation by A. Afanasev {\it et al.}~\cite{Afanasev_DIS}, The solid and dot-dashed curves are calculations by A. Metz {\it et al.}~\cite{Metz}  (multiplied by~$-1$).   {\bf Right panel:}  The average measured asymmetry for the DIS data with combined systematic and statistical uncertainties.}
\label{fig:asym_plot_n}
\end{figure}

The dominant systematic uncertainty for BigBite is from background contamination, the largest of which is from pair-produced electrons, see Table~\ref{table:kine}.  The $\pi^-$ contamination in the T6 triggers  ranges from 0.5 to 2.0\% (rel.) from the lowest to highest $W$ bin, respectively. 
The uncertainties on the contamination are  
$\sim 0.5\%$, which were estimated using the difference between 
information from the Monte Carlo simulation and contamination estimation 
based on data. Further details about these corrections for the 
other two triggers (T1 and T2) can be found in the appendix. 
The uncertainties associated with backgrounds 
contribute to both the asymmetries and dilution factors.  The final results
were extracted taking into account the full correlation of these uncertainties.
Other BigBite systematic uncertainties include the detector acceptance 
(1.2$\times 10^{-4}$), detector response drift (9$\times 10^{-5}$), and 
livetime asymmetry (6$\times 10^{-5}$). For the LHRS, systematic 
uncertainties include the livetime asymmetry (6$\times 10^{-5}$) and 
tracking efficiency (7$\times 10^{-5}$). The correction to the LHRS 
asymmetry due to pair-produced electrons is 1.56 $\times 10^{-4}$ 
with a 100\% relative uncertainty. Systematic uncertainties 
from the polarized target include target polarization and 
misalignment (5\%), and luminosity fluctuations (1.2$\times 10^{-5}$).

The $^3$He and  neutron results are presented in Table~\ref{table:kine} along with the pair-produced electron contamination.  Neutron results are shown in Fig.~\ref{fig:asym_plot_n}. 
 The asymmetry is generally negative and non-zero across the measured kinematic range.  
At the largest value of $W$, the systematic uncertainty is quite large due to the uncertainty in the 
pair-produced electron contamination. In order to evaluate how much the data disfavors the zero-asymmetry hypothesis in the DIS region, the average asymmetry was calculated for the data with
$W>2.0$ GeV.  Because the 
systematic uncertainties of the BigBite points are mostly due to background contamination, they were assumed to be fully correlated, and uncorrelated with the LHRS point.  
The final average neutron asymmetry in the DIS region and its total experimental uncertainty are determined to be $ (-1.09 \pm 0.38) \times10^{-2}$, which is non-zero at the 2.89$\sigma$ level.  
The data are in good agreement with the  two-photon exchange prediction by A.\ Metz {\it{et al.}}~\cite{Metz}, $A_y^n \sim -10^{-2}$,  that uses model input from the semi-inclusive DIS Sivers distribution.      

We have presented the first measurements of the neutron target-normal SSA, $A_y^n$, in the DIS region using a polarized $^3$He target.  Because $A_y$  must be zero at Born-level its measurement is a valuable laboratory for  studying two-photon exchange and the dynamics of the nucleon beyond the simple quark-parton model.  Further measurements  for both proton and neutron with higher precision over a broader kinematic range are necessary to gain a deeper understanding of the role of two-photon exchange in nucleon structure studies.

We acknowledge the outstanding support of the Jefferson Lab
Hall A technical staff and Accelerator Division in accomplishing this experiment. We thank A. Afanasev, C. Weiss and A. Metz for their valuable theoretical guidance.  This work was supported in
part by the U.S. National Science Foundation, the UK Science and Technology Facilities Council,
the U.S. Department of Energy and by DOE contract DE-AC05-06OR23177, under which
Jefferson Science Associates, LLC operates the Thomas
Jefferson National Accelerator Facility.

\nocite{*}
\section{Appendix}   
\noindent
The tables in this appendix show the values used for the corrections in equations~(5) and~(6). 
  Here, we use the notation $f_4 = Y_2/Y_1$, $f_3 = Y_3/Y_2$, and $f_5= C \cdot Y_3/Y_2$, with $C=1.8\pm0.4$.  
The triggers are T1: proportional to the total energy deposited in the electromagnetic calorimeter, 
T2: coincidence between gas Cherenkov and calorimeter energy deposited, T6: same as T1 but with higher discriminator threshold.  
The data from the three triggers were corrected for background and combined.
Here, Y$_1$, Y$_2$, Y$_3$ are yields (events normalized by accumulated charge and detector livetime). The unit is events/$\mu$C.

\begin{table*}[h]
\footnotesize
\centering
\begin{tabular}{|c|c|c|c|c|c|c|c|c|c|}\hline
Bin no. 	& T1 	& T1 stat. rel. 	& T1 sys. rel. 	& T2 		& T2 stat. rel. & T2 sys. rel. 	& T6 		& T6 stat. rel. & T6 sys. rel. 	\\\hline
1		&0.0071 	&0.069 		&1 		&0.0034 	&0.14 		&1 		&0.0038 	&0.031 		&1 			\\
2		&0.019  	&0.031 		&0.6 		&0.008 		&0.053 		&0.6 		&0.0076 	&0.016 		&0.6 			\\
3		&0.044 		&0.013 		&0.35 		&0.014  	&0.027 		&0.35 		&0.013  	&0.0082 	&0.35		\\ 	
4		&0.084  	&0.008 		&0.35 		&0.016  	&0.021 		&0.35 		&0.016  	&0.0062 	&0.35 		\\
5		&0.11 	 	&0.006 		&0.35 		&0.020  	&0.018 		&0.35 		&0.023  	&0.0047 	&0.35 		\\
\hline
\end{tabular}
\centering	
  \caption{Tabulated $f_1$ and its errors.}
\label{table:f1}
\end{table*}

\begin{table*}[h]
\centering
\begin{tabular}{|c|c|c|c|c|c|c|c|c|c|}\hline
Bin no. 	& T1 	& T1 stat. rel. 	& T1 sys. rel. 	& T2 		& T2 stat. rel. 	& T2 sys. rel. 	& T6 		& T6 stat. rel. 	& T6 sys. rel. 	\\\hline
1		&3.90 	&0.0208 	&0 			&2.74 		&0.00894 	&0 			&3.76 		&0.00698 	&0 		\\ 	
2		&6.13 	&0.0166 	&0 			&4.31 		&0.00712 	&0 			&5.89 		&0.00557 	&0 			\\
3		&12.8 	&0.0115 	&0 			&8.73 		&0.00500 	&0 			&12.0 		&0.00389 	&0 			\\
4		&16.3 	&0.0102 	&0 			&9.65 		&0.00475 	&0 			&13.3 		&0.00370 	&0 			\\
5		&32.5 	&0.00721 	&0 			&11.2 		&0.00440 	&0 			&15.7 		&0.00340 	&0 			\\
\hline
\end{tabular}
\centering	
  \caption{Tabulated $Y_1$ and its errors.}
\label{table:N1}
\end{table*}

\begin{table*}[h]
\centering
\begin{tabular}{|c|c|c|c|c|c|c|c|c|c|}\hline
Bin no. 	& T1 		& T1 stat. rel.	& T1 sys. rel. 	& T2 		& T2 stat. rel. 	& T2 sys. rel. 	& T6 		& T6 stat. rel. 	& T6 sys. rel. 	\\\hline
1		&0.0310 	&1.00 	        &0.15 		&0.065	 	&0.123 		&0.15 		&0.0645 		&0.151 		&0.15 		\\ 	
2		&0.232	 	&0.378 	        &0.15 		&0.191 		&0.0718 	&0.15 		&0.259	 		&0.0756 	&0.15 		\\
3		&2.29 		&0.119   	&0.15 		&1.00 		&0.0313 	&0.15 		&1.43 			&0.0321 	&0.15 		\\	
4		&5.64 		&0.0761 	&0.15 		&2.32 		&0.0206 	&0.15 		&3.37 			&0.0209 	&0.15 		\\	
5		&20.9 		&0.0394 	&0.15 		&5.82 		&0.0130 	&0.15 		&8.24 			&0.0134 	&0.15 		\\	
\hline	
\end{tabular}
\centering	
  \caption{Tabulated $Y_2$ and its errors.}
\label{table:N2}
\end{table*}

\begin{table*}[ht]
\centering
\begin{tabular}{|c|c|c|c|c|c|c|c|c|c|}\hline
Bin no. 	& T1 		& T1 stat. rel. 	& T1 sys. rel. 	& T2 		& T2 stat. rel. 	& T2 sys. rel. 	& T6 		& T6 stat. rel. 	& T6 sys. rel.	 \\\hline
1		&0.0310 	&0.185	 	&1 		&0.0154 	&0.0734 	&1 		&0.0239 	&0.0718 	&1 			\\
2		&0.173 		&0.125 		&0.6		&0.0569	 	&0.0394 	&0.6 		&0.0736 	&0.0390 	&0.6 			\\
3		&0.945	 	&0.0506 	&0.35 		&0.224 		&0.0191 	&0.35 		&0.291	 	&0.0185 	&0.35 		\\
4		&2.42 		&0.0298 	&0.35 		&0.312 		&0.0143 	&0.35 		&0.403 		&0.0141 	&0.35 		\\
5		&7.03 		&0.0166 	&0.35 		&0.530	 	&0.00937 	&0.35 		&0.905 		&0.00858 	&0.35 		\\
\hline
\end{tabular}
\centering	
  \caption{Tabulated $Y_3$ and its errors.}
\label{table:N3}
\end{table*}

\begin{table*}[ht]
\centering
\begin{tabular}{|c|c|c|c|c|c|c|c|c|}
\hline
Setup 			& Bin 1 			& Bin 2 		& Bin 3 			& Bin 4 		& Bin 5 			\\\hline
T1 electron 		&	0.0132		& 0.00930	& -0.000216		& 0.0163	& -0.00392		\\
T1 electron abs. err.	&	0.0104		& 0.00837	& 0.00578		& 0.00515	& 0.00369		\\
T2 electron 		&	-0.00429	& 0.00161	& 0.00723		& 0.0000208	& -0.00200		\\
T2 electron abs. err.	&	0.00778		& 0.00619	& 0.00434		& 0.00412	& 0.00381		\\
T6 electron 		&	0.000968	& 0.00524	& 0.00309		& -0.00343	& -0.00574		\\
T6 electron abs. err.	&	0.00259		& 0.00211	& 0.00149		& 0.00145	& 0.00141		\\\hline
T1 positron 		&	-0.0198		& -0.0742	& -0.0449		& -0.0160	& -0.0281		\\
T1 positron abs. err.	&	0.0841		& 0.0405	& 0.0171		& 0.0110	& 0.0064	 	\\
T2 positron 		&	0.00812		& 0.0239	& -0.0372		& -0.0103	& -0.0218		\\
T2 positron abs. err.	&	0.0674		& 0.0343	& 0.0154		& 0.0103	& 0.00709		\\
T6 positron 		&	0.0329		& -0.00966	& -0.0229		& -0.0217	& -0.0188		\\
T6 positron abs. err.	&	0.0228		& 0.01182	& 0.00537		& 0.00364	& 0.00258		\\\hline
T1 $\pi^{-}$ 		&	-0.0402		& -0.0204	& -0.00313		& 0.0251	& 0.0201		\\ 
T1 $\pi^{-}$ abs. err.	&	0.01688		& 0.00921	& 0.00427		& 0.00293	& 0.00189		\\
T2 $\pi^{-}$ 		&	-0.0495		& -0.0479	& 0.0138		& 0.0394	& 0.0237		\\
T2 $\pi^{-}$ abs. err.	&	0.0305		& 0.0175	& 0.00886		& 0.00672	& 0.00490		\\
T6 $\pi^{-}$ 		&	-0.0735		& -0.0391	& -0.00061		& 0.0255	& 0.0185		\\
T6 $\pi^{-}$ abs. err.	&	0.00979		& 0.00565	& 0.00287		& 0.00216	& 0.00157		\\\hline
T1 $\pi^+$ 		&	-0.00313	& 0.0148	& -0.00276		& -0.0185	& -0.00876		\\
T1 $\pi^+$ abs. err.	&	0.0364		& 0.0165	& 0.00709		& 0.00480	& 0.00341		\\
T2 $\pi^+$ 		&	0.0576		& -0.00439	& 0.00625		& -0.0020	& -0.0181		\\
T2 $\pi^+$ abs. err.	&	0.0401		& 0.0218	& 0.0123		& 0.0125	& 0.0150		\\
T6 $\pi^+$ 		&	0.00939		& -0.00534	& -0.0156		& -0.0189	& -0.0165		\\
T6 $\pi^+$ abs. err.	&	0.0131		& 0.00716	& 0.0042		& 0.00428	& 0.00566		\\\hline
\end{tabular}
\centering	
  \caption{Raw asymmetries for each type of particle, corrected for beam charge and livetime.}
\label{table:asym}
\end{table*}

\bibliography{DIS_Ay_PRL_V20}
\end{document}